\documentclass[reprint,amsmath,amssymb,aps,prapplied,floatfix]{revtex4-2}
\makeatletter
\usepackage[hidelinks,colorlinks=true,linkcolor=blue,citecolor=blue, urlcolor = blue]{hyperref}
\usepackage{bookmark}
\newcommand*{\rom}[1]{\expandafter\@slowromancap\romannumeral #1@}
\usepackage{parskip}
\usepackage{graphicx}
\usepackage{dcolumn}
\usepackage{bm}
\usepackage{comment}
\usepackage[normalem]{ulem}
\usepackage[utf8]{inputenc}
\usepackage{hyperref}
\usepackage{lineno}
\usepackage{changepage}
\usepackage{multirow}
\usepackage{array}
\usepackage{float}
\usepackage{amsmath}
\usepackage{titlesec}
\usepackage{rotating}
\usepackage{cleveref}
\usepackage{feynmp-auto}
\usepackage{pgfplots}
\usepackage{graphicx}
\usepackage{adjustbox}
\pgfplotsset{compat=1.15}
\usepackage{mathrsfs}
\usepackage{silence}
\usetikzlibrary{arrows}
\begin{document}
\preprint{APS/123-QED}
\title{Comparative analysis of C-moments using different phenomenological models} 
\author {Jinu James}
\email{jinujames24032001@gmail.com}
\affiliation{Materials Science Centre, Indian Institute of Technology, Kharagpur, West Bengal 721302, India}
\author {R. Aggarwal}
\email{ritu.aggarwal1@gmail.com}
\affiliation{USAR, Guru Gobind Singh Indraprastha University, East Delhi Campus, 110092, India}
\author {M. Kaur}
\email{manjit@pu.ac.in}
\affiliation{Department of Physics, Panjab University, Chandigarh 160014, India\\
Department of Physics, Amity University, Punjab, Mohali 140306, India
}
\date{\today}
\begin{abstract} 
Analysis of normalised C-moments of multiplicity distribution calculated from the different phenomenological models, the Bialas-Praszalowics~(BP) model, modified negative binomial and the superposed SGD at different center of mass~(cms) energies is presented.~The analysis covers a range of energies~(200-900 GeV) of $\overline{p}p$ collisions in restricted phase space slices.~A comparison of different models to the experimental data on charged particle multiplicity spectra from the $\bar{p}p$ annihilation in five pseudorapidity windows is reported.~The comparison shows that the two approaches other than the BP model are in better agreement to the data.~Results on variation of moments with pseudorapidity window size and with center of mass energy and observations from such a study in $\bar{p}p$ annihilation and $pp$ interactions at the same cms energy are also presented.
\end{abstract}
\maketitle
\section{{INTRODUCTION}\label{Int}}
The number of charged hadrons produced in a high-energy-particle collision is one of the basic quantities that are measured extensively for interpreting the interaction dynamics.~The corresponding probability distribution in the phase space under study is known as the multiplicity distribution.~All experiments using different combinations of probes and targets at increasing center-of-mass~(cms) energies in $e^+e^-$, $ep$, $pp$, $p$A and AA collisions, where A stands for ion, aim to measure the multiplicity.~Relating to this, several theoretical, phenomenological models and Monte Carlo generators have been used for validating the predictions on the multiplicity distributions and observing the trends in average number of hadrons produced.                                                                                      
It is well established that the final $n$-charged particles probability distribution,~$P_n$, at lower cms energies and in full phase space is narrower than a Poisson distribution.~With the increase in collision energy, it gradually becomes Poissonian and at higher cms energies, broader than a Poisson distribution.~Independent emission of single particle yields is a Poissonian distribution. Deviations from this shape, therefore reveal correlations.~Higher-order moments and their cumulants are the precise tools to study these correlations between the produced particles.\\\\ 
Around the year 1972, the authors Polyakov~\cite{Polyakov:1970lyy} followed by Koba, Nielsen and Olesen~\cite{KOBA1972317} proposed that at very high collision energies $\sqrt{s}$, the probability distributions $P_n$ should exhibit the scaling~(homogeneity) relation, often called as Koba, Nielsen and Olesen~(KNO) scaling, expressed as;
\begin{equation}
P_{n}=\lim_{s\to\infty} \frac{1}{\langle n_s\rangle}\Psi\left(\frac{n}{\langle n_s\rangle}\right) 
\end{equation}
where $\langle n_s\rangle$ $\equiv$ $\langle n\rangle$ = $\sum_{n=0}^{n_{max}} nP_{n}$ is the average charged multiplicity.~The scaling property was foreseen as an important source of dynamics of hadron production at high energies.~However the KNO scaling violation was observed at energies around 540~GeV.~In order to account for the scaling violations and for a better understanding of the hadron production mechanism, the behaviour could be successfully explained by a negative binomial distribution (NBD).~However, with the availability of data at 900 GeV from the proton-antiproton collisions, the NBD showed pronounced deviation~\cite{ua5charged, alner1985}.~Significant deviations were also observed later on, in the data at higher energies from the Large hadron Collider~(LHC)~\cite{CMSEx}, with the multiplicity distribution showing a shoulder structure.~Several models were developed to understand this structure.~Giovannini et al \cite{Giovannini_1999} discussed that the multiplicity distribution in full phase space is influenced by the energy-momentum and charge conservation, while the distribution in restricted phase space is less affected by such constraints and is expected to be a more sensitive probe of the underlying dynamics.~This can be used as an effective way of constraining  phenomenological models.~Following this, many experiments measured multiplicity in restricted rapidity windows, a practice which is widely accepted.~In a recent publication \cite{Germano} multiplicity moments at the LHC energies have been computed and compared with the predictions of two simple models; Kharzeev-Levin (KL) model \cite{KL1,KL2} and Bialas-Praszalowicz (BP) \cite{BP1,BP2} model.~The focus of the work has been to show that the BP model is able to reproduce the moments for the LHC data at cms energy of 0.9,~2.36 and~13~TeV.~The KL model however fails for some energies, the reasons for the failure are also mentioned.~The multiplicity distribution is represented by the probabilities of $n$-particle events as well as by its moments or its generating function.~Calculated as derivatives of the generating function, the moments and their analysis is a powerful tool for unfolding the characteristics of the distribution.~The multiparticle correlations can be studied through the normalized moments.

The high energy scattering is studied within Quantum Chromodynamics~(QCD).~The approach to these phenomena is to divide them into two broad categories: i) hard or semi-hard involving large momentum exchange producing events with mini-jets which can be computed in a perturbative framework and ii) soft - involving small momentum exchanges producing events without mini-jets in intrinsically non-perturbative process.~The second category of events dominate the bulk of QCD cross-sections.~It was suggested by F. Gelis et al \cite{2010ARNPS} that the traditional separation of hard versus soft QCD dynamics is oversimplified because semi-hard scales generated dynamically at high energies can lead to understanding non-perturbative phenomena in QCD using weak coupling methods.~A new paradigm to compute the scattering dynamics in hadrons and nuclei was introduced as the Color Glass Condensate~(CGC) description. A review and basic concepts of the effective theory for the color glass condensate which describes the high-energy limit of QCD interactions can be found in \cite{Gelis2004bz,2010ARNPS}.~It was emphasised that the CGC could be applied to study a wide range of high energy scattering experiments from Deep Inelastic Scattering (DIS) at HERA and the future Electron Ion Collider~(EIC) to proton/deuterium-nucleus and nucleus-nucleus experiments at the Relativistic Heavy Ion Collider~(RHIC) and LHC.~In the context of multiplicity studies, CGC approach predicted that at very high energies the multiplicity distribution would become narrower.~This implies that the multiplicity moments should decrease with energy. However, the LHC data on the $pp$ collisions at $\sqrt{s}\geq$ 7 TeV did not validate the prediction as the multiplicity distributions are not narrowing down.~In addition, moments' analysis has not been presented by most of the experiments gathering data.

This paper presents a detailed analysis of the data from proton-antiproton annihilation from three different methods for analysis of moments.~By comparing and contrasting our analysis with the BP model, we shall discuss the relative advantages of our procedure.~Although multiplicity distributions have been extensively studied for the data from colliders, not much emphasis has been put to inspect the behaviour of moments of distributions.~In sections~\ref{BPM} and ~\ref{SG} a description of the three approaches is given.~Section~\ref{mmm} gives the definition of the moments in BP model.~Section~\ref{Res} gives the results.~Section~\ref{Unc} describes the calculation of uncertainties on the moments followed by conclusion in section~\ref{Con}.
\section{{BIALAS-PRASZALOWICZ MODEL AND THE NEGATIVE BINOMIAL DISTRIBUTION } \label{BPM}}
It was proposed by A. Bialas~\cite{BP2} that the measured multiplicity distributions can be described as a superposition of Poisson distributions in the form;
\begin{equation}
    P_{n}(BP) = \int_{0}^{\infty}F(x)e^{-\overline{n}x} \frac{({\overline n}x)^n}{n!}dx \label{eq:pnb}
\end{equation}
This formula showed that the function $F(x)$ can be interpreted as the distribution of the amount of matter produced in the collision, for a random emission of particles resulting from its decay, with $x$ as a fraction of the average multiplicity.~Thus $F(x)$ is the distribution of sources that contribute the fraction $x$ to the multiplicity probability $P_{BP}(n)$. 
The normalization condition requires that:
\begin{equation}
    \int_{0}^{\infty}F(x)dx = \int_{0}^{\infty}xF(x)dx= 1  \label{eq:norm}
\end{equation}
The factorial moments of multiplicity distribution measure directly the $F_{m+1}$ moments of the source.
As a result of Eq.(\ref{eq:norm}) $\langle n\rangle$=$\overline{n}$. $F$ is given by
\begin{equation}
    F(x, k) = \frac{k^k}{\Gamma(k)}x^{k-1}e^{-kx} \label {eq:BP}
\end{equation}
where $\Gamma$ is the gamma function and $P_{BP}(n)$ turns out to be a negative binomial distribution (NBD) with parameter $k$.~In fact, $k^{-1}$ can be well approximated by a linear function of \text{ln}$\sqrt{s}$ \cite{Grosse_2010}.
Factorial moments can be expressed through scaled regular moments, C$_q$.

Later on, an explicit form of $F(x)$ top be NBD was adopted in~\cite{BP1} to describe multiplicity moments.~This forms the basis of the BP model. For NBD;
\begin{equation}
F_{m+1} = \frac{k(k+1+.....(k+m)}{k^{m+1}}
\end{equation} 

The  C$_q$ moments then can be calculated starting from;
\begin{equation}
    C_2 = F_2 +\frac{1}{\langle n\rangle}
\end{equation}
The NBD has successfully matched the low energy experimental data on multiplicity observations in different particle-collision experiments, $\overline{p}p$, $hh$, $hA$, $AA$, $e^{+}e^{-}$ ($h$-hadron, $A$-heavy-ion) \cite{TARNOWSKY201351,Cugnon}.~It remained one of the most widely used statistical models for all experimental studies of multiplicity distributions.~However, it faced the first challenge from the $\overline{p}p$ collision data at 540~GeV. A distinct shoulder-like structure in the multiplicity distribution at high energies could not be described by a single NBD but rather required the superposition of two NBDs to model multiplicities.~The idea firstly suggested by C. Fuglesang \cite{Fuglesang:1989st,gio1990multi} and later developed by Giovannini et al~\cite{Giovannini_1999,PRD59,NBD1} is to include the effect of the weighted superposition of two NBDs.~One accounting for the soft events~(events without mini-jets) and second for the semi-hard events~(events with mini-jets).~The contribution of mini-jets grows rapidly with increase in cms energy.~The weight being $\alpha_{sf}$ for the fraction of soft events ($sf$) and $(1-\alpha_{sf})$ for the semi-hard ($sh$).
\begin{equation}
P_n^{Two-NB}=\alpha_{sf} P_n^{NB}(\overline{n}_{sf}, k_{sf})+(1-\alpha_{sf})P_n^{NB}(\overline{n}_{sh}, k_{sh}) \label{eq:2NB} 
\end{equation}
A third NBD was proposed \cite{Zborovský_2013} to describe additional features of the high energy $pp$ data at the Large Hadron Collider~(LHC). However, increasing the number of fit parameters introduces larger errors. The applicability of two-NBD distribution was successful up to the LHC energies of 8~TeV.~Results from the BP model calculations are described in section~\ref{Res}.
\section{{SHIFTED GOMPERTZ DISTRIBUTION}\label{SG}}
~Shifted Gompertz distribution~(SGD) is a distribution of two independent random variables one of which has an exponential distribution and the other has a Gumbel distribution, also known as log-Weibull distribution.~The two non-negative fit parameters $c$ and $\alpha$, where $c>$ 0 define the scale parameter and $\alpha>$ 0 the shape parameter.~The probability distribution function~(PDF) of a random non-negative variable X is given by;
\begin{equation}
P_X(x;c,\alpha) = c e^{-(cx + \alpha e^{-cx})}\big(1+\alpha(1-e^{-cx})\big), \>\>\>\>\>\>\>\>\>x>0
\end{equation}
The normalised moments ( C$_q$) are defined as:
\begin{equation}
C_q = \frac{E[X^{q}]}{(E[X])^{q}} 
\end{equation} 
The distribution has been widely studied in various contexts \cite{Nonlinear, Jimenez:2008:NMC, Torres:2014:EPS}.~In one of our papers \cite{Chawla}, we introduced the shifted Gompertz distribution to investigate the multiplicity distributions of charged particles produced in leptonic and hadronic collisions at the Large Electron-Positron Collider~(LEP), at the Super Proton Synchrotron~(SPS) and collisions at the LHC at different center of mass energies in full and in restricted phase space.~It was shown  that this distribution can be successfully used to study the multiplicities in high energy particle collisions.~In this paper, we include the calculations of higher moments of the multiplicity distribution using superposed SGD.~The two-NBD adopts the two-prong data (soft and semi-hard) approach.~Similar concept is used in SGD.~Many analyses of moments~\cite{PhysRevD54,ZPhys75,PRD96_Panday,PRL106,ALIC} have been done at different energies, using different probability distribution functions and different type of particles. 
\section{{MOMENTS OF MULTIPLICITY DISTRIBUTION}}\label{mmm}
Normalised moments of $q^{th}$-order of the multiplicity distribution are defined as  C$_{q}$ = $\frac{\langle {n^q}\rangle}{\langle {n}\rangle^q}$. 
In the BP model, as described in~\cite{BP1} the moments C$_q$ are obtained from Eq.~(\ref{eq:pnb}) and given by;
\begin{gather}
C_2 = \frac{1}{\langle n\rangle}+1+\frac{1}{k}     \label{eq:C2}\\
C_3 = C_{2}(2C_{2}-1)-\frac{C_{2}-1}{\langle n\rangle}  \label{eq:C3}\\
\begin{split}
C_4 = C_{2}(6C_{2}^2-7C_{2}+2)-2\frac{3C_{2}^2-4C_{2}+1}{\langle n\rangle}\\+\frac{C_{2}-1}{\langle n\rangle^2} \label{eq:C4}
\end{split}
\end{gather}
\begin{equation}
\begin{split}
C_5 = C_{2}(24C_{2}^3-46C_{2}^2+29C_{2}-6)\\-2\frac{18C_{2}^3-34C_{2}^2+19C_{2}-3}{\langle n\rangle}\\+\frac{14C_{2}^2-23C_{2}+9}{\langle n\rangle^2}-\frac{C_{2}-1}{\langle n\rangle^3} \label{eq:C5}
\end{split}
\end{equation}
We analyse the data from the experiment UA5~\cite{ua5charged, alner1985} on $\bar{p}p$ collisions at three energies; $\sqrt{s}$ = 200, 540, 900~GeV in five pseudorapidity windows.~The reason to revisit these data is the availability of precision data in different pseudorapidity windows enabling an extended analysis.~UA5 was the first experiment which observed the violation of KNO scaling in full phase space at 546~GeV and later confirmed it also at 900 GeV.
\section{{RESULTS}\label{Res}}
\subsection{{Moments from BP model}\label{mom}}
Following the method prescribed in \cite{BP1, Germano} the experimental values of $\langle n\rangle$ are fitted using the equations:
\begin{gather}
\langle n\rangle = \sum_{n} nP(n) = exp(\Delta Y) = \left(\frac{1}{x}\right)^\Delta \label{eq:BFKL}\\
 x = \frac{q_{0}^2}{s} \label{eq:x}
\end{gather}
where $q_0$ is a constant, $\Delta$ is the Balitsky-Fadin-Kuraev-Lipatov~(BFKL) intercept parameter \cite{BFKL} and $s$ is the square of cms energy.~From Eqs.~(\ref{eq:BFKL}, \ref{eq:x}) we obtain the values of $\Delta$ and $q_0$ for different data sets.~Further the $\langle n\rangle=\left(\frac{s}{q_{0}^2}\right)^\Delta$ and the moments are obtained from Eq.~(\ref{eq:pnb}).~The moments  C$_2$ are parameterised as;
\begin{equation}
C_{2}=a + b~\text{log}(\sqrt{s})   \label{eq:c2param} 
\end{equation}
where $\sqrt{s}$ is in GeV.~Following the procedure detailed in~\cite{BP1}  C$_2$ parameterisation obtained from Eq.~(\ref{eq:c2param}) is substituted in to Eq.~(\ref{eq:C4}) obtaining C$_4$ as a function of $a$ and $b$.~Then, C$_4$ is adjusted to the data and $a$ and $b$ are determined.~C$_2$, C$_3$ and C$_5$ are then calculated.~The procedure adopted here is the same as done in~\cite{BP1} aiming at the consistency of calculations.
Table~\ref{table:one} gives the values of $q_0$ and $\Delta$ for different rapidity regions of the $\bar{p}p$ at three energies.~Table~\ref{table:BP} shows the normalized moments  C$_2$,~ C$_3$,~ C$_4$ and  C$_5$ calculated from the experimental data and the BP-model for $|\eta|\leq$ 0.5,~1.5,~3.0,~5.0 and all-$\eta$ values.
\begin{table}[t]
\caption{$q_0$ and $\Delta$ from Eqs.~(\ref{eq:BFKL},~\ref{eq:x}).~Parameters $a$ and $b$ are obtained from fitting of Eq.~(\ref{eq:C4}) after substituting C$_2$ from Eq.~(\ref{eq:c2param})} \label{table:one}
\begin{tabular}{c c c c c}
\hline\hline
$\eta$ & $q_0$     & $\Delta$   & $a$       & $b$  \\ \hline
0.5    & 6.12      & 0.13       & 1.03      & 0.13  \\ 
1.5    & 0.02      & 0.11       & 1.14      & 0.07  \\ 
3.0    & 0.002     & 0.12       & 1.03      & 0.06   \\ 
5.0    & 0.008     & 0.15       & 1.02      & 0.05   \\ 
all    & 0.030     & 0.12       & 1.06      & 0.03   \\ \hline
\hline
\end{tabular}
\end{table}
\begin{table*}[t]
\caption{Normalised C$_q$ moments calculated from the experimental data \cite{ua5charged, alner1985} and the BP model for different pseudorapidity intervals at different center of mass energies $\sqrt{s}$.} \label{table:BP}
\centering
\begin{tabular}{c c | c c c c c| c c c c c}
\hline\hline
&&   \multicolumn{5}{c|}{} & \multicolumn{5}{c}{}\\
&&\multicolumn{5}{c|}{Moments-Experiment}&\multicolumn{5}{c}{Moments-BP Model}\\ 
& &\multicolumn{5}{c|}{}&\multicolumn{5}{c}{}\\ \hline
 &	&    &     & 	& &  & & & &  &    \\	  
$\sqrt{s}$ (GeV) & $|\eta|$ & $\langle n\rangle$ &  C$_2$  &  C$_3$ & C$_4$ &  C$_5$ & $\langle n\rangle$ &  C$_2$ &  C$_3$ &  C$_4$ &  C$_5$ \\
 &  	 &	  &     &	&	&		  &	  &   &   &   & \\\hline 
&0.5 & 2.47  $\pm$  0.09	& 1.87  $\pm$  0.05	& 4.52  $\pm$  0.25	& 12.94  $\pm$   1.13  & 41.59  $\pm$   5.04 	& 2.42 & 1.87 & 4.76  & 15.34   & 59.57 \\
&1.5	& 7.92  $\pm$  0.25 & 1.54  $\pm$   0.04  &	 3.06  $\pm$  0.15   &  7.19  $\pm$  0.54   &  18.96  $\pm$  1.92   & 7.79 &1.54  & 3.13  & 7.90  & 23.72 \\ 
200& 3.0	& 15.43  $\pm$  0.38  &	1.38  $\pm$  0.03	 & 2.40  $\pm$  0.09	    &  4.90  $\pm$  0.28 &  11.24  $\pm$  0.84 & 15.25 &  1.38 &  2.43 &  5.13  &  12.67 \\
&5.0	& 20.64  $\pm$  0.47 &	1.27  $\pm$  0.02 & 1.95  $\pm$  0.07	 & 3.49  $\pm$  0.18 & 6.98  $\pm$  0.49 & 20.22 & 1.28 &  1.96 &  3.54 &  7.29\\ 
&all	& 21.29  $\pm$  0.68 &	1.24  $\pm$  0.03		 &	1.85  $\pm$  0.08     & 3.12  $\pm$  0.21   & 5.83  $\pm$  0.54  & 20.83 &  1.24 & 1.84  & 3.16  & 6.15 \\\\

&0.5	 & 3.00  $\pm$  0.06&	1.91  $\pm$  0.03 & 4.97  $\pm$  0.16  &  16.13  $\pm$  0.9 & 61.68  $\pm$  5.11  & 3.11 & 1.91  &  5.09 &  17.41  &  72.52 \\
&1.5	& 9.46  $\pm$  0.17  &	1.60  $\pm$  0.02 & 3.45  $\pm$  0.10	 & 9.15  $\pm$  0.43    & 28.37  $\pm$  1.88   & 9.75 &  1.59  &  3.45  &  9.38  & 30.66 \\
540& 3.0	 &18.96  $\pm$  0.31 &	1.47  $\pm$  0.02	 & 2.85  $\pm$  0.07   & 6.76  $\pm$  0.26   & 18.82  $\pm$  1.00  & 19.37 &  1.45 & 2.75  & 6.41  & 17.67 \\
&5.0	& 26.32  $\pm$  0.43  & 1.35  $\pm$  0.02	&	 2.26  $\pm$  0.05   & 4.44  $\pm$  0.16   & 9.83  $\pm$  0.49  & 27.23 & 1.33 & 2.20  & 4.35 & 9.99  \\
&all	 & 28.36  $\pm$  0.43 &	1.30  $\pm$  0.02 & 2.07  $\pm$  0.05   & 3.83  $\pm$  0.14   & 7.98  $\pm$  0.39  & 29.31 & 1.28 & 1.99 & 3.65  & 7.67 \\\\

&0.5	 & 3.61  $\pm$  0.11 &	1.93  $\pm$  0.05	  & 5.39  $\pm$  0.32	  & 19.71  $\pm$  2.17   & 88.98  $\pm$  15.59  & 3.54 & 1.93   &  5.25  &  18.48  &  79.53 \\
&1.5	& 11.11  $\pm$  0.25 &	1.63  $\pm$  0.03	 & 3.55  $\pm$  0.15	     & 9.37  $\pm$  0.50   & 28.12  $\pm$  1.99  & 10.93 &  1.63  &  3.62  & 10.21 & 34.73 \\
900&3.0	 & 22.15  $\pm$  0.36 &	1.48  $\pm$  0.02		 &	2.86  $\pm$  0.06  & 6.74  $\pm$  0.21   & 16.92  $\pm$  0.69  & 21.91 &  1.49  &  2.92  & 7.13 & 20.74 \\
&5.0	& 32.26  $\pm$  0.46 &	1.35  $\pm$  0.01		 &	2.27  $\pm$  0.09     & 4.46  $\pm$  0.11  &   9.73  $\pm$  0.3 & 31.74 &  1.36 &  2.33  & 4.82 &  11.63\\
&all	& 35.47  $\pm$  0.77 &	1.29  $\pm$  0.02		 & 2.04  $\pm$  0.06	  & 3.72  $\pm$  0.17  & 7.52  $\pm$  0.46 & 34.94 &  1.30  &  2.07  & 3.92 & 8.54\\\hline\hline
\end{tabular}
\end{table*}
\subsection{Moments from two-NBD}
\begin{figure}
\includegraphics[scale=0.56]{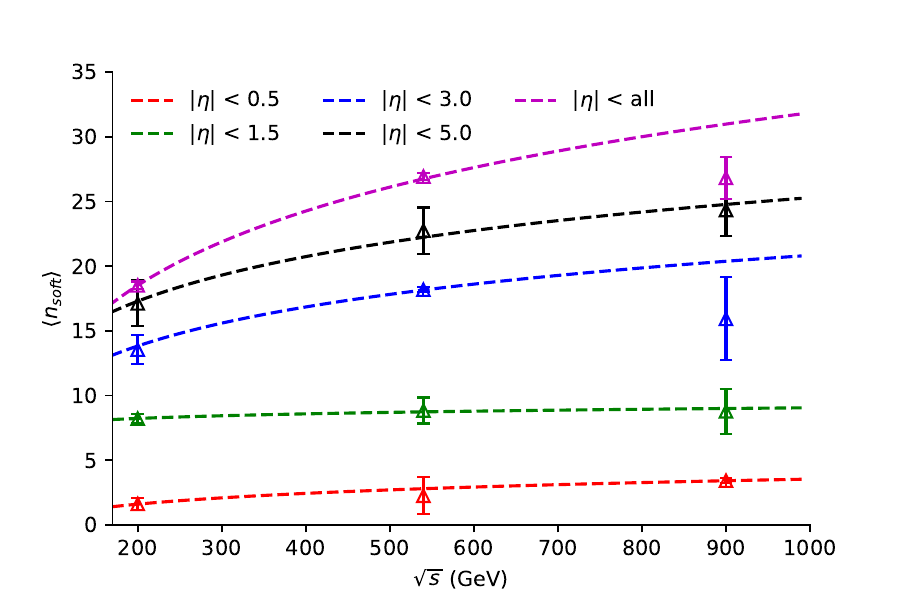}
\includegraphics[scale=0.56]{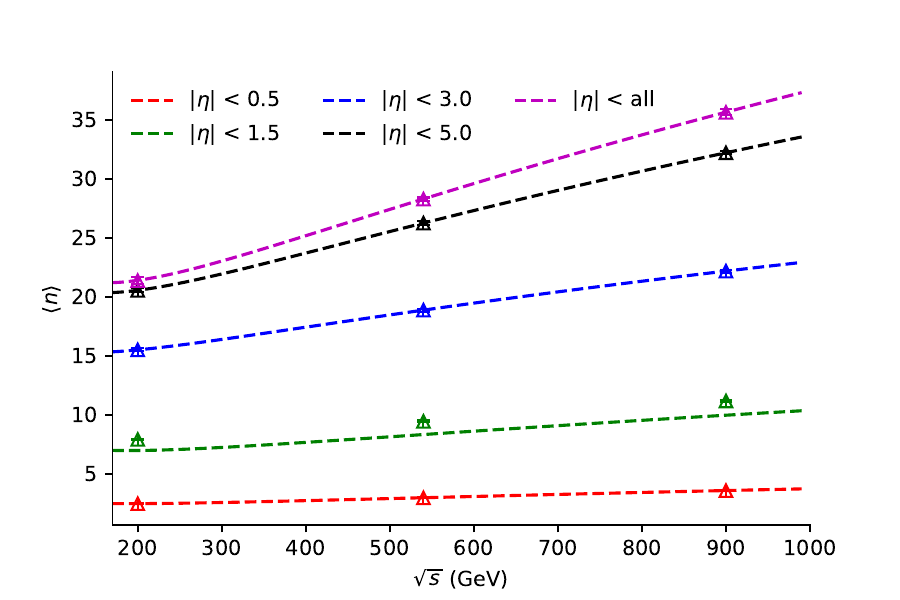}
\caption{Variation of average multiplicity of charged hadrons in soft interactions~$\langle n_{\text {soft}}\rangle$~(top) and in all $\bar{p}p$ interactions $\langle n\rangle$~(bottom) with $\sqrt{s}$. The points represent the values calculated from the two-NBD fits and the dotted lines show the fits with Eqs.~(\ref{eq:Nsoft}) and Eq.~(\ref{eq:cqpar}) respectively for the two plots.} \label{fig:AVN1-N}
\end{figure}

The charged particle multiplicity distribution in full phase space becomes broader than a Poisson distribution at high energies.~The most widely adopted, NBD \cite{VANHOVE} to describe the multiplicity spectra, fails to reproduce the experimental data.~This lead to the introduction of a two-component approach by A. Giovannini et al \cite{NBD1}.
In the two-NBD approach, data are fitted with Eq.~(\ref{eq:2NB}).~Corresponding to the best-fit value of $\alpha_{sf}$ the $P_n$ versus $n$ distribution is obtained which represents the weighted superposition of two NBDs, one representing soft events and second the semi-hard events. 
Table~\ref{Tab:ratio} shows the average multiplicities $\langle{n_1}\rangle$ and $\langle{n_2}\rangle$ corresponding to the two-NBD superposed to obtain the best fit. The ratio of $\frac{\langle{n_2}\rangle}{\langle{n_1}\rangle}$ is obtained.~It is observed that in the full phase space, the average multiplicity of semi-hard interactions is approximately twice the average multiplicity of the soft interactions.~A logarithmic dependence of multiplicity of soft events on cms energy, also reported in~\cite{Giovannini_1999}, is observed as;
\begin{equation}
\langle n_\text{soft}\rangle = a + b~\text{ln}(\sqrt{s}) \label{eq:Nsoft}
\end{equation}
The relation holds for all pseudorapidity intervals, as can be observed from figure~\ref{fig:AVN1-N}
\begin{table}
\caption{Average multiplicity of charged hadrons in soft interactions $\langle n_{1}\rangle$ and semi-hard interactions $\langle n_{2}\rangle$ obtained from two-NBD fits and the ratio of the two.} \label{Tab:ratio}
\begin{tabular}{c c  c c c c c}
\hline\hline
$\sqrt{s}$(GeV)  & $|\eta|$  & $\langle n_1\rangle$ & $\langle n_2\rangle$ & $\frac{\langle n_2\rangle}{\langle n_1\rangle}$\\\hline
 
      & 0.5 & 2.15  $\pm$  0.17  & 3.93  $\pm$  0.58  & 1.83  $\pm$  0.41\\
      & 1.5 & 8.23  $\pm$  0.34  & 8.07  $\pm$  0.47  & 0.98  $\pm$  0.10 \\                    
200   & 3.0 & 13.55  $\pm$  1.12  & 19.21  $\pm$  1.34  & 1.42 $\pm $0.22 \\                  
      & 5.0 & 17.12  $\pm$  1.78  & 29.21 $\pm$  3.10  & 1.71$ \pm$ 0.36 \\                      
      & all & 18.51  $\pm$  0.31  & 35.26  $\pm$  1.25 & 1.91  $\pm$  0.10 \\\\
      & 0.5 & 2.26  $\pm$  1.43  & 4.50  $\pm$  1.42  & 1.99  $\pm$  1.90 \\
      & 1.5 & 8.850  $\pm$  1.01 & 10.16  $\pm$  0.10 & 1.15  $\pm$  0.24 \\                    
540   & 3.0 & 18.18  $\pm$  0.21 & 29.62  $\pm$  1.88 & 1.63  $\pm$  0.12\\                     
      & 5.0 & 22.74  $\pm$  1.79 & 35.62  $\pm$  4.63 & 1.57  $\pm$  0.33\\                     
      & all & 26.92  $\pm$  0.29 & 50.65  $\pm$  3.34  & 1.88  $\pm$  0.15\\\\ 
      
      & 0.5 & 3.43  $\pm$  0.18 & 5.63  $\pm$  1.27 & 1.64  $\pm$  0.45\\
      & 1.5 & 8.79  $\pm$  1.73 & 23.24  $\pm$  6.78 & 2.64  $\pm$  1.29\\                       
900   & 3.0 & 15.93  $\pm$  3.21 & 39.24  $\pm$  10.72 & 2.46  $\pm$  1.17\\                     
      & 5.0 & 24.34  $\pm$  2.01 & 56.14  $\pm$  5.68 & 2.31  $\pm$  0.42\\                      
      & all & 26.82  $\pm$  1.61 & 57.73  $\pm$  4.69 & 2.15  $\pm$  0.30\\ \hline\hline                               
\end{tabular}
\end{table}
\begin{figure}
\includegraphics[scale=0.56]{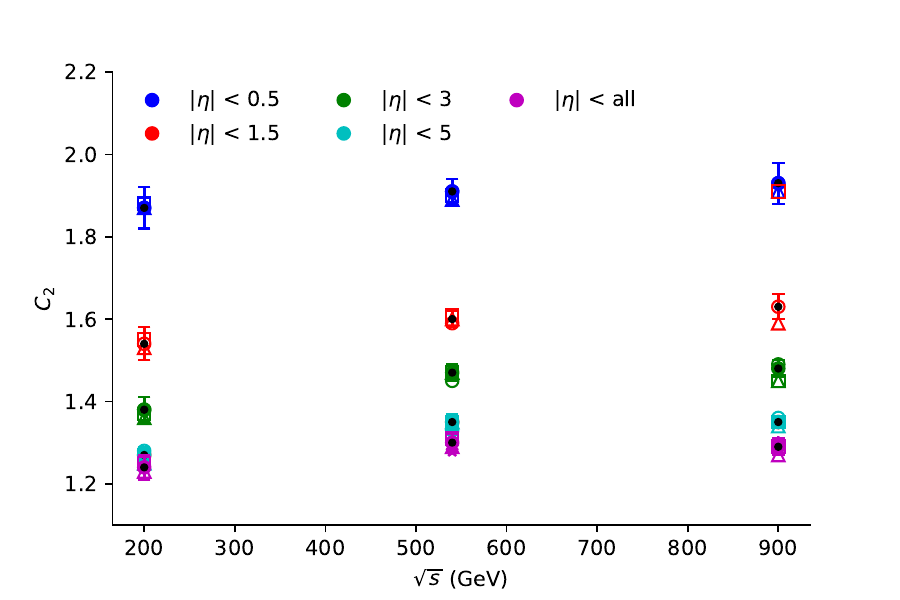} 
\includegraphics[scale=0.56]{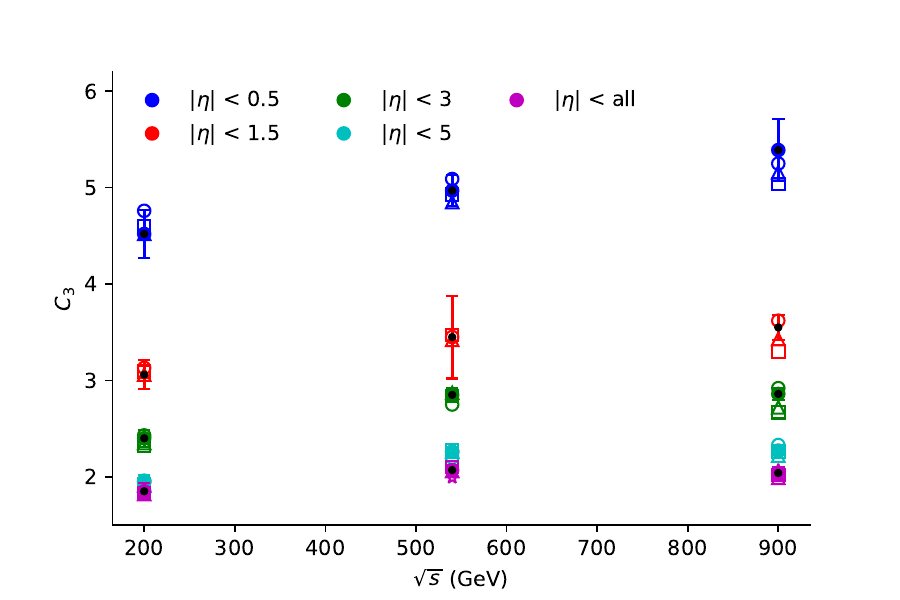} 
\includegraphics[scale=0.56]{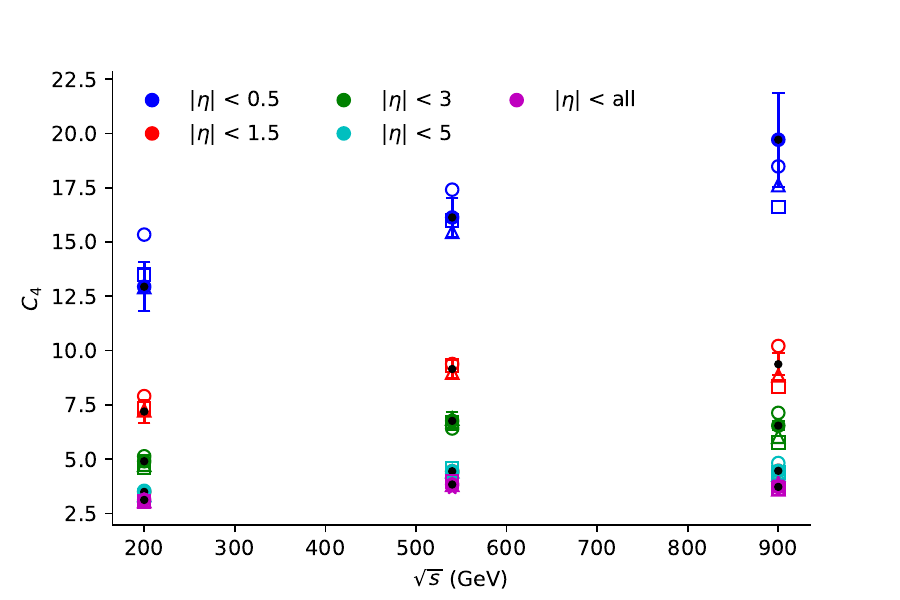} 
\includegraphics[scale=0.56]{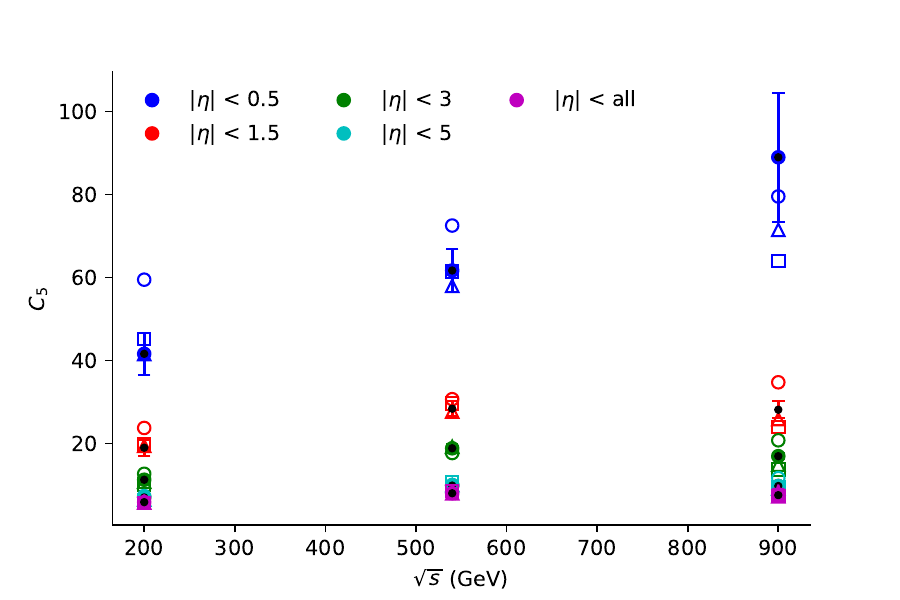} 
\caption{Moments C$_2$,C$_3$,C$_4$,C$_5$~(top to bottom) versus $\sqrt s$ for different $|\eta|$ windows for the $\bar{p}p$ data~($\bullet$), BP model~($\bigcirc$), SGD ($\square$) and two-NBD fit~($\triangle$).}
\label{fig:C2C5}
\end{figure}
\begin{figure}
\includegraphics[scale=0.56]{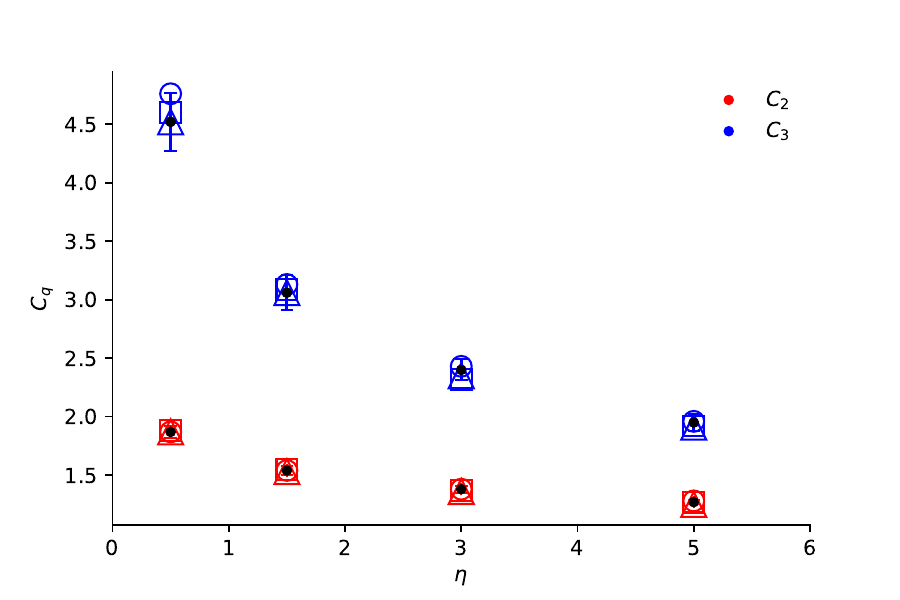} 
\includegraphics[scale=0.56]{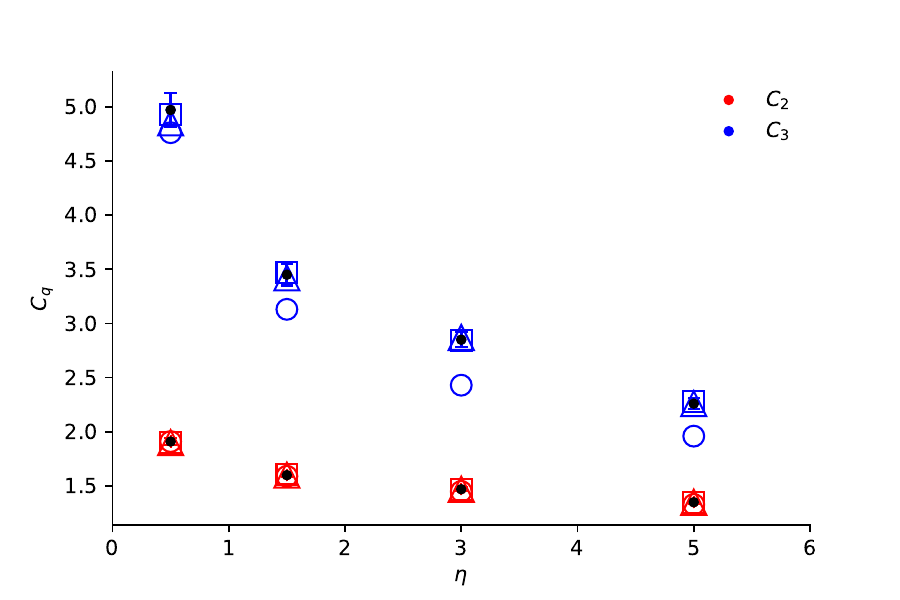}
\includegraphics[scale=0.56]{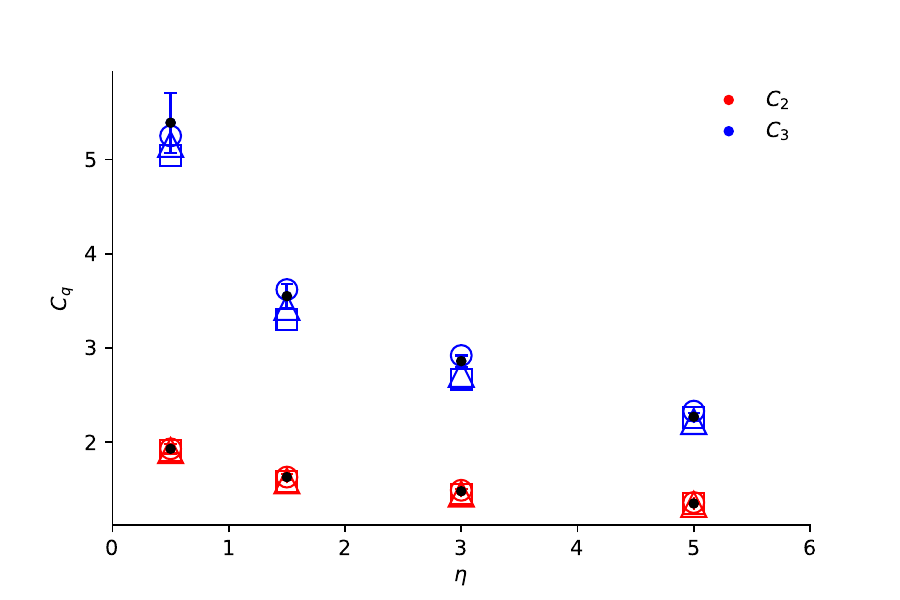}
\caption{Moments C$_{2}$, C$_{3}$ versus $|\eta|$ for $\sqrt{s}$=200, 540, 900~GeV~(top to bottom)for different pseudorapidity $\eta$ for the $\bar{p}p$ data~($\bullet$), BP model~($\bigcirc$), SGD~($\square$) and two-NBD fit~($\triangle$).}
\label{fig:C2C3200-900}
\end{figure}
\begin{figure}
\includegraphics[scale=0.56]{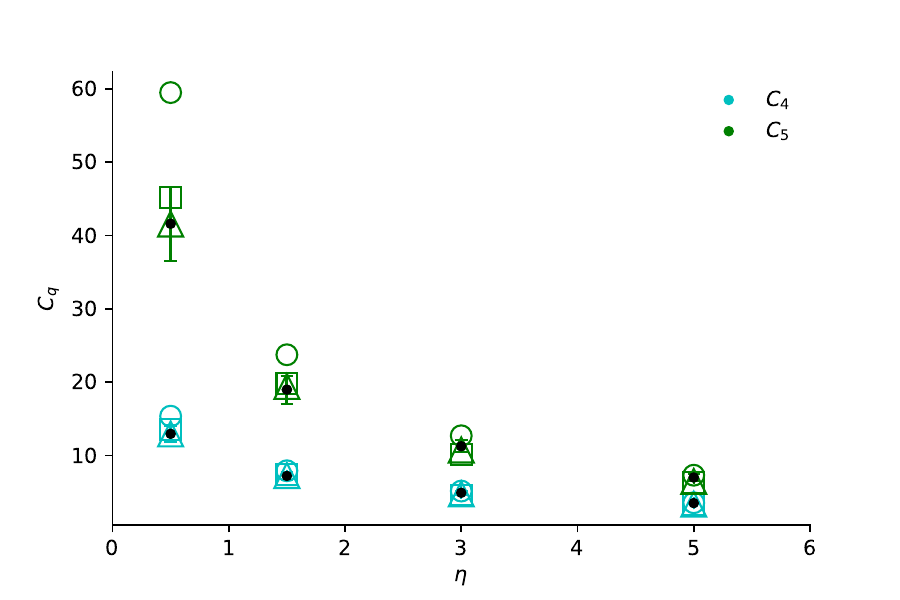}
\includegraphics[scale=0.56]{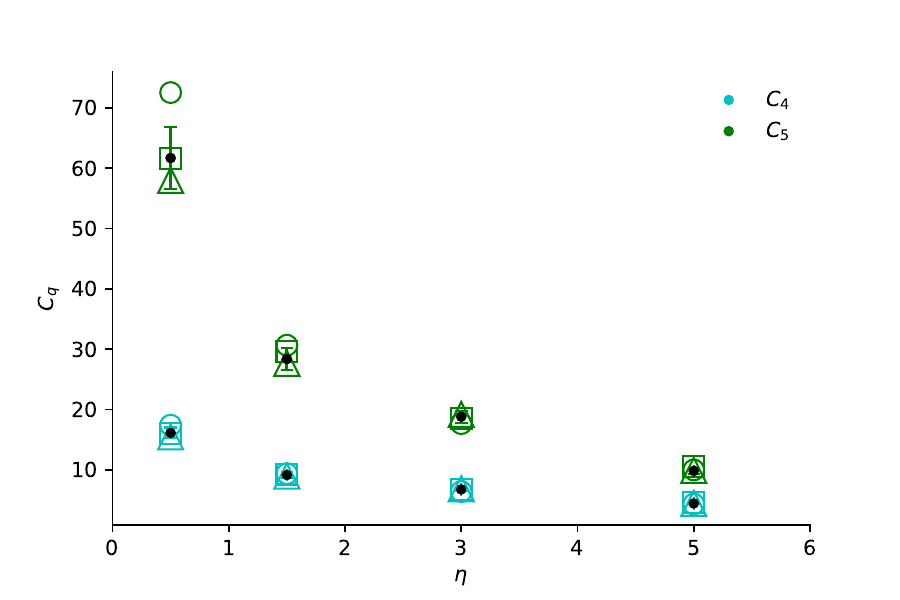}
\includegraphics[scale=0.56]{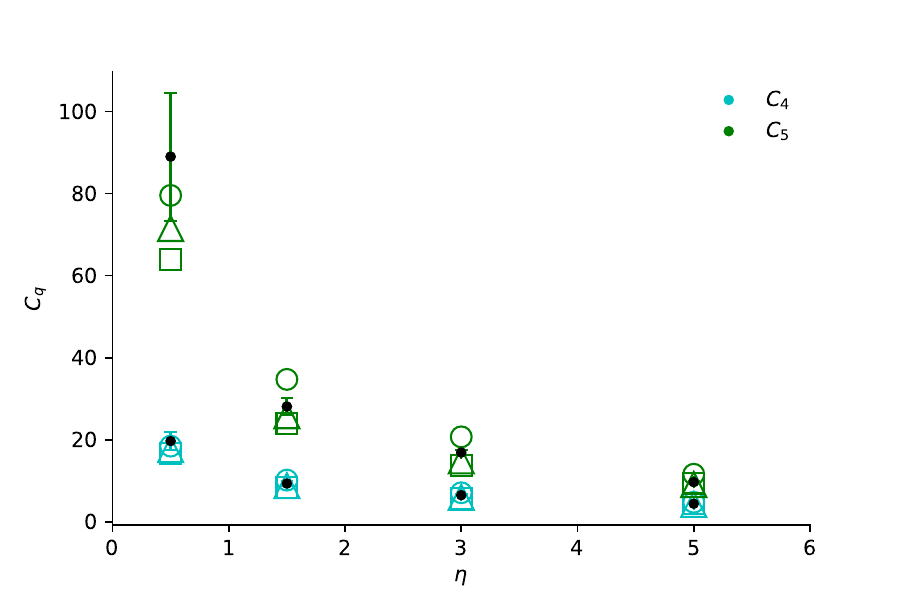}
\caption{Moments C$_{4}$, C$_{5}$ versus $|\eta|$ for $\sqrt{s}$=200, 540, 900~GeV(top to bottom) for different $\eta$ windows for the $\bar{p}p$ data~($\bullet$), BP model~($\bigcirc$), SGD~($\square$) and two-NBD fit~($\triangle$).}
\label{fig:C4C5200-900}
\end{figure}
\begin{figure}
\includegraphics[scale=0.56]{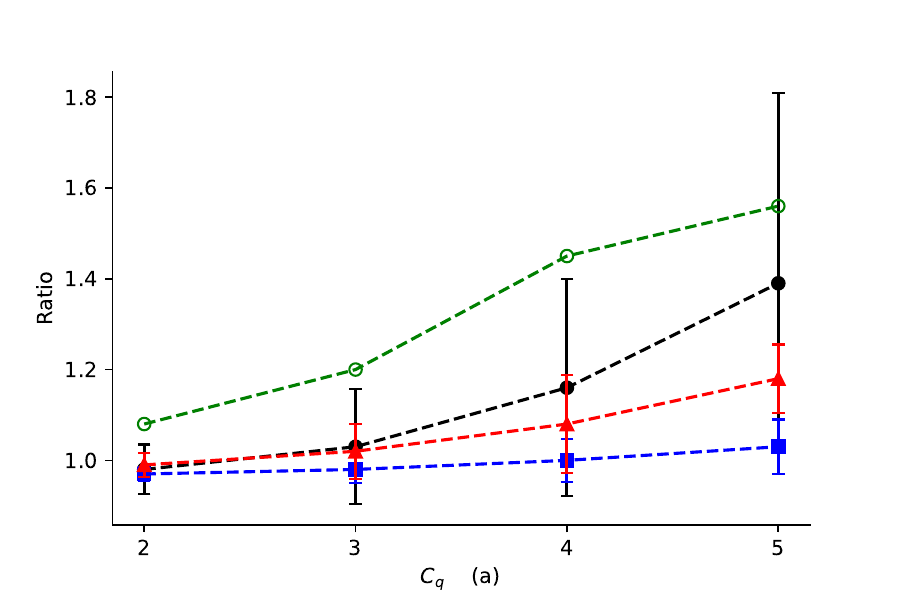}
\includegraphics[scale = 0.56]{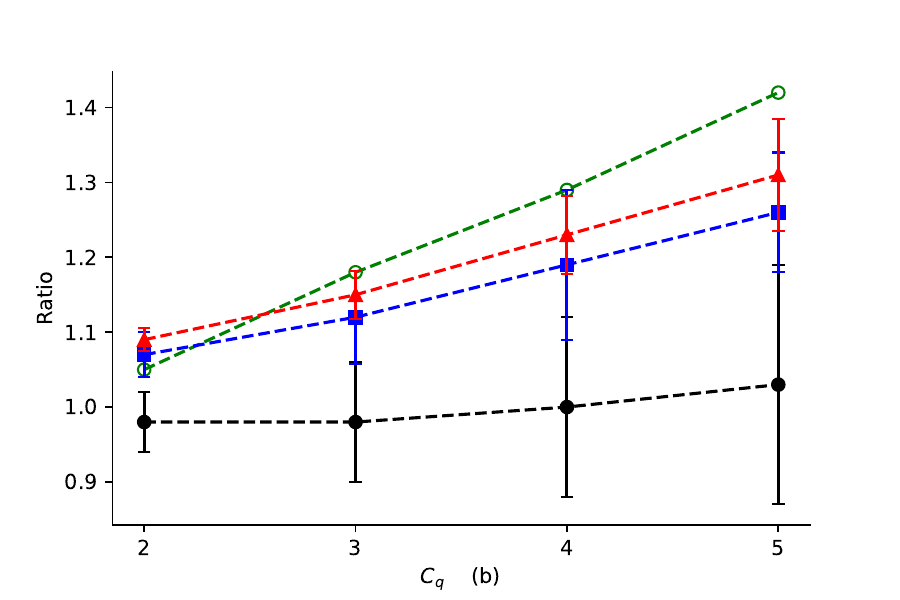}
\caption{Ratio of normalized moments C$_q$($\Bar{p}p/pp$) (a) $\eta < 0.5$ and (b) $\eta < 1.5$~ for the 900 GeV data~($\bullet$) with  two-NBD fit~($\blacktriangle$), SGD~($\blacksquare$) and BP model~($\bigcirc$).}
    \label{fig:TwoEta}
\end{figure}
\begin{figure}
\includegraphics[scale = 0.56]{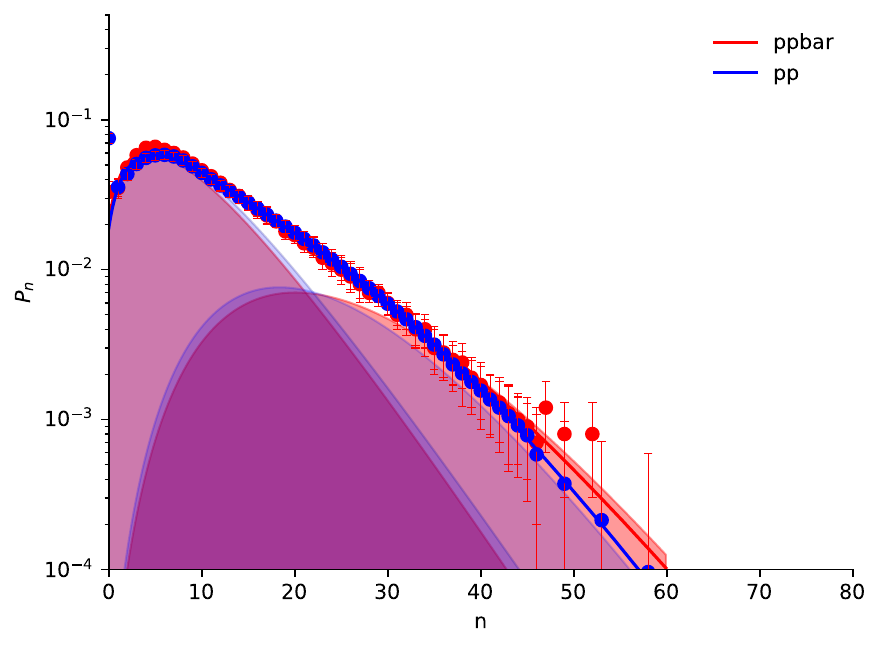}
\caption{Two-NBD fits to $P_n$ versus $n$ distributions of $\bar{p}p$ and $pp$ data for ~$\eta < 1.5$ pseudorapidity showing the soft~(left shaded distribution) and semi-hard components~(right shaded distribution).} \label{fig:frac}
\end{figure}
\begin{table*}
\caption{Normalised C$_q$ moments obtained from two-NBD fit for the $\overline{p}p$ interactions.}\label{table:2nbd3}
{\
\begin{adjustbox}{max width=\textwidth}
\begin{tabular}{c c| c c c c c}
\hline \hline
&  & &\multicolumn{4}{c}{}\\ 
 & & &\multicolumn{4}{c}{Moments~(Two-NBD)}\\\cline{3-7}            
 $\sqrt{s}$ &  &  &\multicolumn{4}{c}{}\\ 
(GeV)& $|\eta|$& $\langle n\rangle$ &C$_2$ & C$_3$ & C$_4$ & C$_5$\\\hline 
& 0.5 & 2.47 $\pm$ 0.06 & 1.87 $\pm$ 0.03  & 4.52 $\pm$ 0.15 & 12.92 $\pm$ 0.66 &41.55 $\pm$ 0.66\\
&1.5	& 7.92 $\pm$ 0.01  & 1.53 $\pm$ 0.06& 3.06 $\pm$ 0.06 &7.25 $\pm$ 0.21 &19.40 $\pm$ 0.77\\
200& 3.0 & 15.51 $\pm$ 0.13 & 1.36 $\pm$ 0.01& 2.35 $\pm$ 0.03& 4.74 $\pm$ 0.08 &10.73 $\pm$ 0.25\\
&5.0	& 20.56 $\pm$ 0.16 &1.25 $\pm$ 0.01& 1.91 $\pm$ 0.02 & 3.34 $\pm$ 0.06 &6.44 $\pm$ 0.06\\
&all	 & 21.39 $\pm$ 0.29 & 1.23 $\pm$ 0.01 & 1.82 $\pm$ 0.04& 3.07 $\pm$ 0.09 &5.74 $\pm$ 0.23\\\\

&0.5	&2.97 $\pm$ 0.03  & 1.89 $\pm$ 0.02 &4.85 $\pm$ 0.08 & 15.47 $\pm$ 0.42& 57.96 $\pm$ 2.18\\
&1.5	 & 9.45 $\pm$ 1.42 & 1.60 $\pm$ 0.01 & 3.42 $\pm$ 0.04 & 8.98 $\pm$ 0.16 & 27.66 $\pm$ 0.68\\
540 &3.0	 & 18.88 $\pm$ 0.12 & 1.47 $\pm$ 0.01 &2.87 $\pm$ 0.09 & 6.87 $\pm$ 0.09 & 19.19 $\pm$ 0.36\\
&5.0  & 26.26 $\pm$ 0.18 & 1.35 $\pm$ 0.01 &2.26 $\pm$ 0.02 &4.44 $\pm$  0.07 & 9.92 $\pm$ 0.07\\
&all	 & 28.29 $\pm$ 0.14 & 1.29 $\pm$ 0.01 & 2.06 $\pm$ 0.02 & 3.82 $\pm$ 0.04 & 7.93 $\pm$ 0.13\\\\

& 0.5	  & 3.58 $\pm$ 0.06 & 1.91 $\pm$ 0.02 & 5.16 $\pm$ 0.14 & 17.62 $\pm$ 0.93 & 71.47 $\pm$ 6.70\\
& 1.5	 & 11.18 $\pm$ 0.10 &1.59 $\pm$ 0.01 & 3.43 $\pm$ 0.04 & 8.86 $\pm$ 0.16 & 25.88 $\pm$ 0.63\\
900& 3.0	 & 22.19 $\pm$ 0.13  & 1.45 $\pm$ 0.01& 2.72 $\pm$ 0.06& 6.03 $\pm$ 0.08 & 14.90 $\pm$ 0.19\\
& 5.0	& 32.22 $\pm$ 0.13 & 1.34 $\pm$ 0.01 & 2.22 $\pm$ 0.01 & 4.29 $\pm$ 0.03 & 9.13 $\pm$ 0.08\\
& all & 35.65 $\pm$ 0.2 & 1.27 $\pm$  0.01 & 1.99 $\pm$ 0.02 & 3.61 $\pm$ 0.06& 7.25 $\pm$ 0.16\\\hline \hline
\end{tabular}
\end{adjustbox}
}
\end{table*}
\subsection{ Moments from shifted Gompertz distribution}
 In an approach very similar to the two-NBD, we use a superposition of two SGDs.~The multiplicity distribution is produced by adding a weighted superposition of multiplicity in soft events and multiplicity distribution in semi-hard events.~It is well understood~\cite{Giovannini_1999,PRD59,NBD1} that this approach combines only the two classes of events and not two different particle-production mechanisms.~Therefore, no interference terms need to be introduced.~The final distribution is the superposition of the two independent distributions.~We call it 'modified shifted Gompertz distribution', the PDF for which is obtained as;
\small
\begin{equation}
P_{n}(\beta:c_1,\alpha_1;c_2,\alpha_2)=\beta P_{n}(\text{soft}) + (1-\beta)P_{n}(\text{semi\textrm{-}hard}) 
\end{equation}
\normalsize
where $\beta$ is the fraction of soft events, (c$_1$, $\alpha_1$) and ( C$_2$, $\alpha_2$) are respectively the scale and shape parameters of the two distributions. Table~\ref{table:SG} shows the values of average multiplicity $\langle n\rangle$, and the moments  C$_q$ for $q$=2-5 in different pseudorapidity intervals and in full phase space for data at $\sqrt{s}$= 200,~540,~900 GeV.
\begin{table*}[t]
\caption{Normalized C$_q$ moments of modified SGD for $\overline{p}p$ collisions.}  \label{table:SG}
{
\begin{adjustbox}{max width=\textwidth}
\begin{tabular}{c c|c c c c c}
\hline\hline
&  & &\multicolumn{4}{c}{} \\
& & &\multicolumn{4}{c}{Moments (Modified shifted Gompertz)} \\\cline{3-7}             
 $\sqrt{s}$&  &  &\multicolumn{4}{c}{}\\
(GeV)& $|\eta|$ &$\langle n\rangle$ &C$_2$ & C$_3$ & C$_4$ & C$_5$     \\\hline
& 0.5 & 2.48  $\pm$  0.04 & 1.88  $\pm$  0.03 & 4.60  $\pm$  0.17 & 13.49  $\pm$  0.74 & 45.17  $\pm$  3.38  \\\ 
& 1.5 & 7.90  $\pm$  0.07 & 1.55  $\pm$  0.02 & 3.09  $\pm$  0.06 & 7.34  $\pm$  0.23 & 19.82  $\pm$  0.82  \\\ 
200 & 3 & 15.54  $\pm$  0.11 & 1.37  $\pm$  0.01 & 2.32  $\pm$  0.03 & 4.59  $\pm$  0.10 & 10.13  $\pm$  0.30  \\\   
& 5 & 20.48  $\pm$  0.13 & 1.27  $\pm$  0.01 & 1.92  $\pm$  0.03 & 3.32  $\pm$  0.07 & 6.29  $\pm$  0.17  \\\  
& all &21.12  $\pm$  0.19 & 1.25  $\pm$  0.01 & 1.83  $\pm$  0.04 & 3.05  $\pm$  0.09 & 5.61  $\pm$  0.22  \\\\

& 0.5 &2.98  $\pm$  0.02 & 1.90  $\pm$  0.02 & 4.93  $\pm$  0.09 & 15.99  $\pm$  0.42 & 61.45  $\pm$  2.19 \\\  
& 1.5 &9.46  $\pm$  0.06 & 1.61  $\pm$  0.01 & 3.47  $\pm$  0.05 & 9.31  $\pm$  0.19 & 29.59  $\pm$  0.83 \\\  
540 & 3 & 18.86  $\pm$  0.09 & 1.47  $\pm$  0.01 & 2.84  $\pm$  0.03 & 6.71  $\pm$  0.11 & 18.57  $\pm$  0.42 \\\    
& 5 &26.179  $\pm$  0.11 & 1.35  $\pm$  0.01 & 2.28  $\pm$  0.02 & 4.58  $\pm$  0.07 & 10.57  $\pm$  0.21 \\\   
& all &28.14  $\pm$  0.10 & 1.31  $\pm$  0.01 & 2.10  $\pm$  0.02 & 3.96  $\pm$  0.05 & 8.55  $\pm$  0.14\\\\

& 0.5 &3.58  $\pm$  0.04 & 1.91  $\pm$  0.02 & 5.04  $\pm$  0.12 & 16.62  $\pm$  0.63 & 63.99  $\pm$  3.46  \\\    
& 1.5 &11.43  $\pm$  0.07 & 1.58  $\pm$  0.01 & 3.30  $\pm$  0.04 & 8.35  $\pm$  0.17 & 23.92  $\pm$  0.68  \\\    
900& 3 &22.03  $\pm$  0.10 & 1.45  $\pm$  0.01 & 2.67  $\pm$  0.03 & 5.76  $\pm$  0.08 & 13.74  $\pm$  0.27 \\\    
& 5 &32.34  $\pm$  0.13 & 1.35  $\pm$  0.01 & 2.26  $\pm$  0.02 & 4.39  $\pm$  0.06 & 9.48  $\pm$  0.16 \\\   
& all &35.73  $\pm$  0.20 & 1.29  $\pm$  0.01 & 2.02  $\pm$  0.02 & 3.65  $\pm$  0.07 & 7.35  $\pm$  0.18 \\\hline\hline 
\end{tabular}
\end{adjustbox}
}
\end{table*}
\subsection{Comparison}
Tables~\ref{table:BP},~\ref{table:2nbd3},~\ref{table:SG} list all values of normalised moments  C$_q$ for $q$=2-5 and the for different $|\eta|$ values for the data and from the three approaches; the BP model, two-NBD and modified SGD.~Figure~\ref{fig:C2C5} shows the comparison of moments  C$_q$ versus $\sqrt{s}$ for different $\eta$ windows for the experimental data, and the three approaches.~From the tables and the figures, it is observed that both two-NBD and modified SGD reproduce the  data very closely for all moments, while the BP model overestimates the data for higher moments.~The moments C$_q$ increase slowly with cms energy $\sqrt{s}$ for lower $q$=2-3, while ramping up quickly for higher values of $q$=4-5.~Figures~\ref{fig:C2C3200-900}-\ref{fig:C4C5200-900} show the variation of all moments with pseudorapidity.~It is observed that moments decrease with increase in pseudorapity $|\eta|$ at all energies and higher moments die down rapidly than the lower moments.

~For two-NBD fit to $P_n$ versus $n$ distribution, table~\ref{Tab:ratio} shows the average multiplicities of the soft interactions~$\langle n_1 \rangle$ and semi-hard interactions~$\langle n_2 \rangle$. It may be observed that in full phase space the ratio of $\frac{\langle n_2\rangle}{\langle n_1 \rangle}$ is $\approx$ 2.~The two plots in figure \ref{fig:AVN1-N} show variation of average multiplicity of charged hadrons in soft interactions ~$\langle n_1 \rangle$=~$\langle n_{\text {soft}}\rangle$~(top) and the charged hadrons in all $\bar{p}p$ interactions $\langle n\rangle$~(bottom) with $\sqrt{s}$ calculated from two-NBD fits.~It is observed that the variation of $\langle n_{\text {soft}}\rangle$ in the upper plot follows the Eq.(\ref{eq:Nsoft}) and $\langle n\rangle$ follows a quadratic rise with logarithm of cms energy as shown in the figure and represented by the following equation;
\begin{equation}
\textrm{C}_{q}=a + b~\textrm{ln}\sqrt{s}+ c~\textrm{ln}^{2}\sqrt{s} \label{eq:cqpar} 
\end{equation}
We also compare the moments of characteristic collision (annihilation) properties of~$\bar{p}p$ with $pp$~(inelastic), the two different type of collisions at the same cms energy, in two common pseudorapidity windows.~The second moment variance captures how spread out a distribution is or its scale parameter.~High variance implies a wider distribution.~While the third moment called skewness, measures the relative size of the two tails of a distribution, the fourth moment called kurtosis is a measure of the combined size of the tails relative to whole distribution.~For this analysis, different $\eta$ regions are constituent parts of the total data samples of 6839 events for $\bar{p}p$ and 132294 events for $pp$ interactions.~Figure~\ref{fig:TwoEta}(a) shows the ratio of moments C$_q$($\bar{p}p/pp$) for $\eta < 0.5$~ at 900 GeV cms energy \cite{ua5charged,CMSEx} for the data, two-NBD, the SGD fit and for the BP model.~This phase space region is dominant in soft events.~Figure~\ref{fig:TwoEta}(b) shows the similar plot for $\eta < 1.5$.~This region has large contribution from semi-hard events.~It is observed that for $\eta <0.5$ the data show a steeper rise in the ratio particularly for moments with  $q >$ 3.~BP model shows the same behaviour but with much higher ratio at each $C_q$.~Both the NBD and superposed SGD do not show such steeper increase.~For $\eta <1.5$, the phase space region has a greater contribution from semi-hard events having mini-jets.~The data shows a ratio closer to one for almost all the moments.~However, all the three models show a steady rise in the ratio.~It is not feasible to study the effects in more pseudorapidity windows due to lack of data.~It is pertinent to mention that the large statistical errors on the $\bar{p}p$ data due to sample size, may influence the comparative observations. 

Further we also compare, each of the models predictions with the data w.r.t the phase space window size.~It is observed that the ratio grows almost linearly with moments, though the data shows a much quicker rise in case of smaller $\eta < 0.5$ than the larger phase space $\eta < 1.5$.~The ratio C$_q$($\Bar{p}p/pp$) for $\eta < 0.5$ is always much larger than unity, while it stays close to unity for $\eta < 1.5$.~Thus the contribution of particles coming from events without mini-jets~(soft) and events with mini-jets~(semi-hard) interactions and the dynamics of particle production in the two type of interactions are different.~It is also observed that while the ratios in case of two-NBD and SGD are in agreement with the data within the limits of errors, the disagreement with the BP model is larger, particularly for one pseudorapidity interval.~However, this result pertains to only two pseudorapidity windows since similar data for comparison in other pseudorapidity windows are not available for $pp$ interactions.

To understand the dynamical differences between $\bar{p}p$ and $pp$ processes, figure~\ref{fig:frac} shows the two-NBD fits to $P_n$ versus $n$ distributions of the two processes for ~$\eta < 1.5$ showing the soft events and semi-hard events as two components.~The reason for choosing ~$\eta < 1.5$ window is due to the observation of KNO scaling violation.~In small ~$\eta$ intervals KNO scaling holds being a property of the multiplicity distribution in cascade process of single jet with self similar branchings and leading order coupling \cite{cohen,DOKSHITZER,Wrochna}.~The left shaded distributions in the figure enclose the distributions for soft events and the right shaded distributions show the semi-hard events for both $\bar{p}p$ and $pp$ interactions.~Small differences between the two type of interactions are observable.~The $\bar{p}p$ has larger multiplicity tail, indicative of the embedded multi-component structure of final state which may be deeper than the corresponding $pp$ case.
\section{UNCERTAINTIES ON MOMENTS}\label{Unc}
A normalised probability distribution where $P_{n}$ is the probability of producing $n$ particles with an uncertainty $\epsilon_{n}$, and assuming that individual bin errors are uncorrelated, the errors on moments can be calculated by using the partial derivatives, as described in \cite{Grosse_2010};
\begin{equation} 
\frac{\partial C_q}{\partial p_n} = \frac{n^{q}\langle n\rangle - \langle n^{q}\rangle qn}{\langle n \rangle ^{q+1}}
\end{equation}
The total error is then
\begin{equation}
E_{q}^2 = \sum_{n}\Big(\frac{\partial C_{q}}{\partial P_n}\epsilon_n \Big)^2 
\end{equation} 
In this analysis, the errors on multiplicities have been taken as the quadratic sum of statistical and systematic errors.~While fitting the data, the data points at the tail end of the probability distribution, which have statistical and systematic errors $>$ 50$\%$ have been dropped and all calculations are done devoid of these.
\section{{CONCLUSION}\label{Con}}
The analysis of C-moments shows a good agreement of the normalized moments derived from the two-NBD approach and from the SGD with the experimental values obtained from the data on $\bar{p}p$ collisions at $\sqrt{s}$ = 200,~540,~900 GeV.~Moments from the BP model also approximate the experimental values within errors albeit the higher moments, C$_4$ and  C$_5$ overestimate the data.~In large pseudorapidity intervals, like in full phase space, the C-moments increase with energy rapidly, while they are roughly constant up to $q$=3 in small pseudorapidity interval, $|\eta<0.5$.~The strong almost linear increase of moments with cms energy indicates a clear violation of scaling.~For small $\eta$ intervals KNO scaling holds being a property of single jet with self similar branchings and leading order coupling.~In the weighted superposition of two-NBD, where one NBD represents the soft events with average multiplicity $\langle n_1\rangle$ and the second NBD represents the semi-hard events with average multiplicity $\langle n_2\rangle$.~The ratio $\frac{\langle n_{2}\rangle}{\langle n_{1}\rangle}$ is $\approx$~2 in full phase space.~Results on moments from the experimental data are better reproduced in both two-NBD and SGDs than the BP model.~A comparison of the moments of $\bar{p}p$ (annihilation process) and $pp$ collisions (inelastic collisions) at the same cms energy of 900 GeV for each of the three models separately shows that the ratio grows almost linearly with moments, though it a much quicker rise in case of smaller $\eta < 0.5$ than in the larger phase space $\eta < 1.5$.~Further, it is observed that while the ratios in case of two-NBD and SGDs are in agreement with the data within the limits of errors, the BP model disagrees, particularly for one pseudorapidity interval.~Thus the contribution of particles coming from events without mini-jets (soft) and events with mini-jets (semi-hard) interactions is different in the two type of interactions.~Small differences between $\bar{p}p$ and $pp$ interactions are observed in the particle multiplicities from the semi-hard events with mini-jets.~This shows that the dynamics of contribution of particles coming from soft and semi-hard interactions may be different in the two type of interactions.
\section*{ACKNOWLEDGEMENTS}
Author Jinu James acknowledges the Ministry of Education, Govt. of India for the Post Graduate Fellowship.
\bibliography{JN.bib}
\end{document}